\documentclass[final]{svjour2}
\usepackage{graphicx}
\usepackage{rotating}
\usepackage{amssymb}
\usepackage{mathptmx}
\usepackage[numbers]{natbib}
\usepackage[outdir=./]{epstopdf}
\usepackage{mathtools}
\newcommand{\rb}{{\bf r}}
\makeatletter
\journalname{Journal of Low Temperature Physics}

\bibpunct{}{}{,}{s}{}{,}

\begin{document}

\newcommand{\hdblarrow}{H\makebox[0.9ex][l]{$\downdownarrows$}-}
\title{Vortex nucleation through fragmentation in a stirred resonant Bose-Einstein condensate}

\author{M.C. Tsatsos$^1$ \and A.U.J. Lode$^2$}

\institute{1:Instituto de F\' isica de S\~ao Carlos, Universidade de S\~ao Paulo, Caixa Postal 369, 13560-970 S\~ao Carlos, S\~ao Paulo, Brazil.\\
\email{marios@ifsc.usp.br}
\\2: Department of Physics, University of Basel, Klingelbergstrasse 82, CH-4056 Basel, Switzerland.}

\date{01.10.2014}

\maketitle

\keywords{Ultracold Bose gas, quantized vortex, many-body physics, MCTDHB, http://ultracold.org}

\begin{abstract}

Superfluids are distinguished from ordinary fluids by the quantized manner the rotation is manifested in them. Precisely, quantized vortices are known to appear in the bulk of a superfluid when subject to external rotation. In this work we study a trapped ultracold Bose gas of $N=101$ atoms in two spatial dimensions that is stirred by a rotating beam. We use the multiconfigurational Hartree method for bosons, that extends the mainstream mean-field theory, to calculate the dynamics of the gas in real time. As the gas is rotated the wavefunction of the system changes symmetry and topology. We see a series of resonant rotations as the stirring frequency is increased. Fragmentation accompanies the resonances and change of symmetry of the wavefunction of the gas. We conclude that fragmentation of the gas appears hand-in-hand with resonant absorption of energy and angular momentum from the external agent of rotation.

\end{abstract}

\section{Introduction}
In sharp contrast to ordinary viscous fluids, supefluids absorb energy and angular momentum from a thread moving inside the superfluid only beyond some critical velocity (see Refs.\cite{PethickSmith, Butts1999, Winiecki1999, Neely2010, Khamis2013} and references therein). For instance, a rotating potential in an atomic Bose-Einstein condensate is known to lead to a series of critical frequencies beyond which nucleation of quantized vortices is favourable\cite{Butts1999, Jackson1998, Tsubota2002, Lundh2003}. Interestingly, when vortices are nucleated the system is driven from a state of zero to some state of finite angular momentum. Thus, the time evolved wavefunction is of different geometry and topology\cite{Dagnino2009}. Furthermore, much recent work has been devoted in exploring the role of fragmentation in attractive\cite{Tsatsos2014} and repulsive\cite{Oksana2014} parabolic traps in different number of dimensions. 
However, the connection between vortex nucleation and fragmentation and, thus, loss of coherence throughout the dynamics has just begun being explored\cite{Weiner2014}. In this paper, by computing  the fragmentation as a function of time, we numerically study the time evolution of the repulsive Bose gas that is initially in a coherent, fully condensed state and stirred with a rotating potential. Here and hereafter, we use the term `condensed' to refer to the states where all particles occupy the same single-particle state and `fragmented' for those that occupy macroscopically a finite number of single-particle states or orbitals.
In order to explore phenomena as fragmentation one needs to go beyond the standard single-orbital mean-field approach like the Gross-Pitaevskii theory. A theory that allows for a good self-consistent description of fragmented states is the multiconfigurational time-dependent Hartree method for bosons (MCTDHB)\cite{RoleOfExcited,MCTDHB} that is briefly discussed below. By applying the MCTDHB method to a two-dimensional gas of $N=101$ repulsive ultracold bosons we demonstrate how this transition to states of different symmetry and topology, such as the quantum vortex states, are accompanied and assisted by fragmentation of the gas. 

\section{Methods \label{methods}}
We model the problem of a stirred gas of ultracold ($T=0$) bosons in two spatial dimensions with the time-dependent many-body Hamiltonian of the form:
\begin{equation}
H(t) = T + V(t) + W, 
\label{Hamiltonian0}
\end{equation}
where the many-body kinetic $T$, potential $V$ and interaction $W$ energy operators respectively take on the forms:
\begin{eqnarray}
\label{Hamiltonian}
T = -\frac 1 2 \sum_i^N \nabla^2_{\rb_i}, ~~ V = \sum_i^N \left[V_0(\rb_i) + V_{\text{rot}}(\rb_i,t)\right], ~~~~~~ \\
\label{potential}
V_0(r)=\frac{\omega_{\text{trap}}^2}{2} \rb^2 ~~\text{and} ~~ V_{\text{rot}}(\rb,t) = A \exp\left\lbrace-\left[(x-x_0(t))^2+(y-y_0(t))^2\right]/\sigma^2 \right\rbrace,  \label{potentials} ~~~~~~ \\
 ~~ W = g_0\sum_{i<j} \frac{1}{2\pi\sigma_{int}^2}\exp[-(\rb_i-\rb_j)^2/(2 \sigma_{int}^2)], ~~~~~~ \label{intOperator}
\end{eqnarray}
where all quantities are expressed in dimesionless units, which we obtain by dividing the Hamiltonian $H$ by $\hbar^2/(m L^2)$, where $L$ is an arbitrary length scale and $m$ is the boson mass. 
The function $V_{\text{rot}}$ represents a laser beam of Gaussian shape, initially at point $(x_0,y_0)$ that rotates with constant angular frequency $\Omega$ as $x_0(t)=\cos(\Omega t), y_0(t)=\sin(\Omega t)$. Such a Gaussian potential is a common way to represent a focused blue detuned laser beam from the atomic resonance. The particles are assumed to interact with the short range two-body interaction potential that is modelled with a Gaussian-shaped function [see Eq.~\eqref{intOperator}]. 

We represent the solution with the many-body ansatz
\begin{equation}
\Psi(t) = \sum_k C_k(t) \Phi_k(t),
\label{MBAnsatz}
\end{equation}
where the time-dependent functions $\{\Phi_k\}$ form an orthonormal set of wavefunctions, each one of which represents a condensed \emph{or} fragmented Bose gas of $N$ particles. Each $\Phi_k$ is a permanent i.e., fully symmetrized product of orbitals. The sum runs over all possible $\binom{N+M-1}{N}$ configurations $\Phi_k=\Phi_k(r_1,r_2,\dots,r_N; t)$ of $N$ particles in $M$ orbitals. The method used to solve the time-dependent Schr\"odinger equation is the MCTDHB and has been presented in Refs.\cite{RoleOfExcited,MCTDHB}. This method takes into consideration $M$ different single-particle states (orbitals) for the construction of the many-body state $\Psi$ and is in principle and in practice numerically exact\cite{Lode2012, Lode2014}. The numerical package and latest recursive implementation (R-MCTDHB), used here to solve the corresponding equations together with a detailed description, can be found in Ref.\cite{Lode2014, ultracoldweb}.

In the present work we attack the rotating problem with the R-MCTDHB numerical package using $M=3$ orbitals and $N=101$ particles. The numerical simulations are done on a $128^2$ grid. The frequency of the trapping potential is $\omega_{\text{trap}}=1$. The value of the interaction strength is fixed to $g=g_0(N-1)=50$. We choose the amplitude for the Gaussian rotating laser beam equal to $A=72$ and width $\sigma=0.01$. The width of the Gaussian function that models the two-body interaction interaction is $\sigma_{int}=0.25$. This is well in the parameter range where the physics of the true contact interaction are emulated by a short-range Gaussian, see Ref.\cite{Doganov2013}.

\section{Results}
In this section we present the main results obtained by solving the time-dependent Schr\"odinger equation
\begin{equation}
i \partial_t \Psi(t) = H \Psi(t),
\label{SE}
\end{equation}
with the above described Hamiltonian, for various stirring frequencies $\Omega$. Within MCTDHB we propagate the quantum state $\Psi(t)$ in real time for the potential of Eq.~\eqref{potential} and $0.5\leq \Omega< 1.5$, expressed here and hereafter in units of the confining frequency $\omega$ and calculate the total energy of the system $E(t)=\langle \Psi(t)| H(t)|\Psi(t)\rangle$. The initial state $\Psi(t=0)$ is the eigenstate of the many-body Hamiltonian of Eq.~\eqref{Hamiltonian0} with $V_{\text{rot}}\equiv 0$ in Eqs.~\eqref{potentials}. We find the time-dependent (natural) orbitals $\phi_k(\vec{r};t),~k=1,2,3$, their respective (natural) occupations $n_k(t)$ and the total density of the gas $\rho(\vec{r};t)=\sum_k^3 n_k(t) |\phi_k(\vec{r};t)|^2$. Last, the fragmentation entropy is defined as (cf. also Ref.\cite{Brezinova2012})
\begin{equation}
\label{fragmentation}
f(t) = -\sum_k^M \frac{n_k(t)}{N} log_2 \frac{n_k(t)}{N}
\end{equation}
and gives information on the distribution of the occupations and hence how many different configurations $\Phi_k$ [see Eq.~\eqref{MBAnsatz}] contribute to the many-body dynamics. Thus, a fully condensed system in some state $\phi_m$ with $n_m=N$ and $n_j=0$ for $j\neq m$ gives $f=0$ while the maximum value $f_{max}=-log_2(1/M)$ corresponds to the maximally fragmented system with $n_1=n_2=\dots=n_M=N/M$. In our case, $M=3$ and $f_{max}\approx 1.58$.

\begin{figure}
\begin{center}
\includegraphics[width=0.80\linewidth,keepaspectratio]{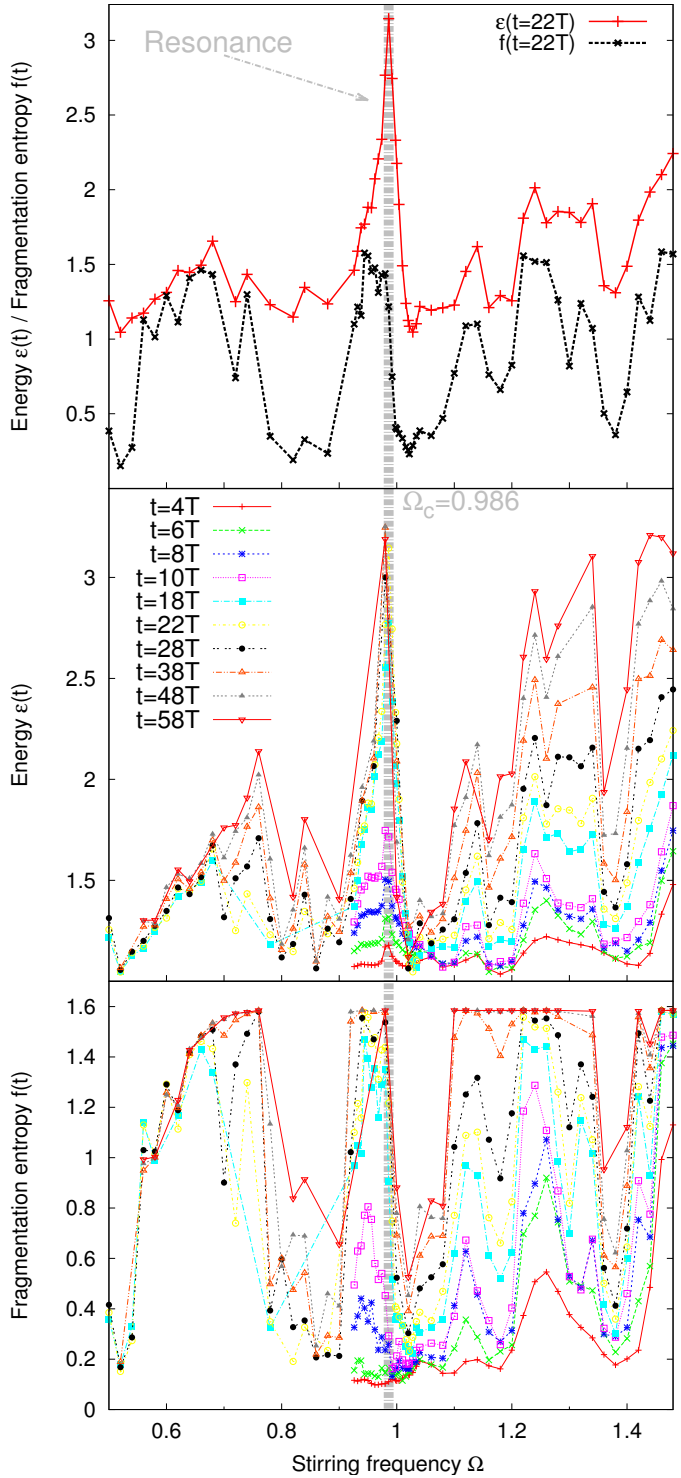}
\end{center}
\caption{(Color online) Resonances in the dynamics of the 2D Bose gas with the stirring frequency $\Omega$. The red (solid) line in the upper panel shows the scaled total energy of the ground state $\epsilon(t)=E(t)/E(0)$ as a function of the stirring frequency $\Omega$. The most profound resonance appears around the value $\Omega_c=0.986$ (gray vertical line throughout all panels), where the energy is absorbed the fastest. The resonances here are for time $t=22T$ (where $T$ is the stirring period), but they appear even at earlier times (see middle panel). At later times the principal resonance moves towards smaller values of $\Omega$. The black (dashed) line in the upper panel is the fragmentation entropy $f$ (see also main text). Interestingly, the fragmentation entropy $f$ as a function of the stirring frequency follows the same pattern as the energy $\epsilon(t)$. In the middle panel we plot the scaled energies $\epsilon(t)$ as functions of $\Omega$ for various times ($t=4,6,8,10,18,22,28,38,48,58$ stirring periods $T$). The resonances appear at practically the same frequencies for all different examined values of stirring time. In the lower panel we plot the total fragmentation entropy $f$ at the same times. Practically, after $t=38T$ $f$ reaches its maximum value $f_{\text{max}}$ for most of the $\Omega$ values examined, leaving thus ``windows'' of $\Omega$ with off-resonant, low-f states.}
\label{Fig_EnergiesOm}
\end{figure}

\paragraph{Resonances with $\Omega$.} 
We plot the resulting $\epsilon(t) = E(t)/E(0)$ and $f(t)$, at times $t=n T, n\in\mathbb N$, i.e., integer multiples of the stirring periods, in Fig.~\ref{Fig_EnergiesOm}. The system shows a principal resonance around a value $\Omega_c$, which at $t=22T$ we found to be $\Omega_c=0.986$. For later times the resonance moves to slightly smaller values of $\Omega$. Also, several secondary resonances appear at, mainly, larger frequencies. At $\Omega=0.986$ the system absorbs the available energy from the external rotor in the fastest way, at least for times as high as $t=48T$, where $T$ is the period of the Gaussian-shaped stirrer, described in Sec.~\ref{methods}. In a non-interacting rotor the system would resonate exactly at $\Omega=1=\omega_\text{trap}$ (see Ref.\cite{Dagnino2009}) and the energy absorbed would increase unbounded and exponentially in time. We see a remarkable correlation of the energy with the fragmentation entropy $f$ of the system, as can be seen in Fig.~\ref{Fig_EnergiesOm}.

The energy pumped into the system from the environment (rotor) excites different modes, depending on $\Omega$. As we shall see, centre-of-mass rotation, expansion of the cloud and vortices appear simultaneously or separated. In the most profound resonance ($\Omega=0.986$) all three types of excitations coexist. To make this statement more transparent we now examine distinct representative cases that show the different types of excitations in the densities.

\paragraph{Resonant stirring without vortex nucleation.}
The stirring of the gas and the energy that flows into the system cause a variety of excitations, not necessarily vortex-like excitations. 

For instance, at $\Omega=1$ only centre-of-mass rotation is seen for stirring times as high as $t=64T$. Fragmentation increases very slowly, after $65$ stirring periods the first orbital is still $80\%$ occupied and the fragmentation entropy is at about $f=0.9$. Interestingly, the total energy of the system oscillates in time with a very long period, that appears to be close to $55$ stirring periods.
Moreover, at the value $\Omega=1.14$ we also observed no vortex formation. Despite the gas being on resonance (the energy doubles and the fragmentation increases to its maximum $f_{max}$ at $t=40T$) the net result of the stirring is the expansion of the size of the gas.

\begin{figure}
\begin{center}
\includegraphics[width=0.75\linewidth,keepaspectratio]{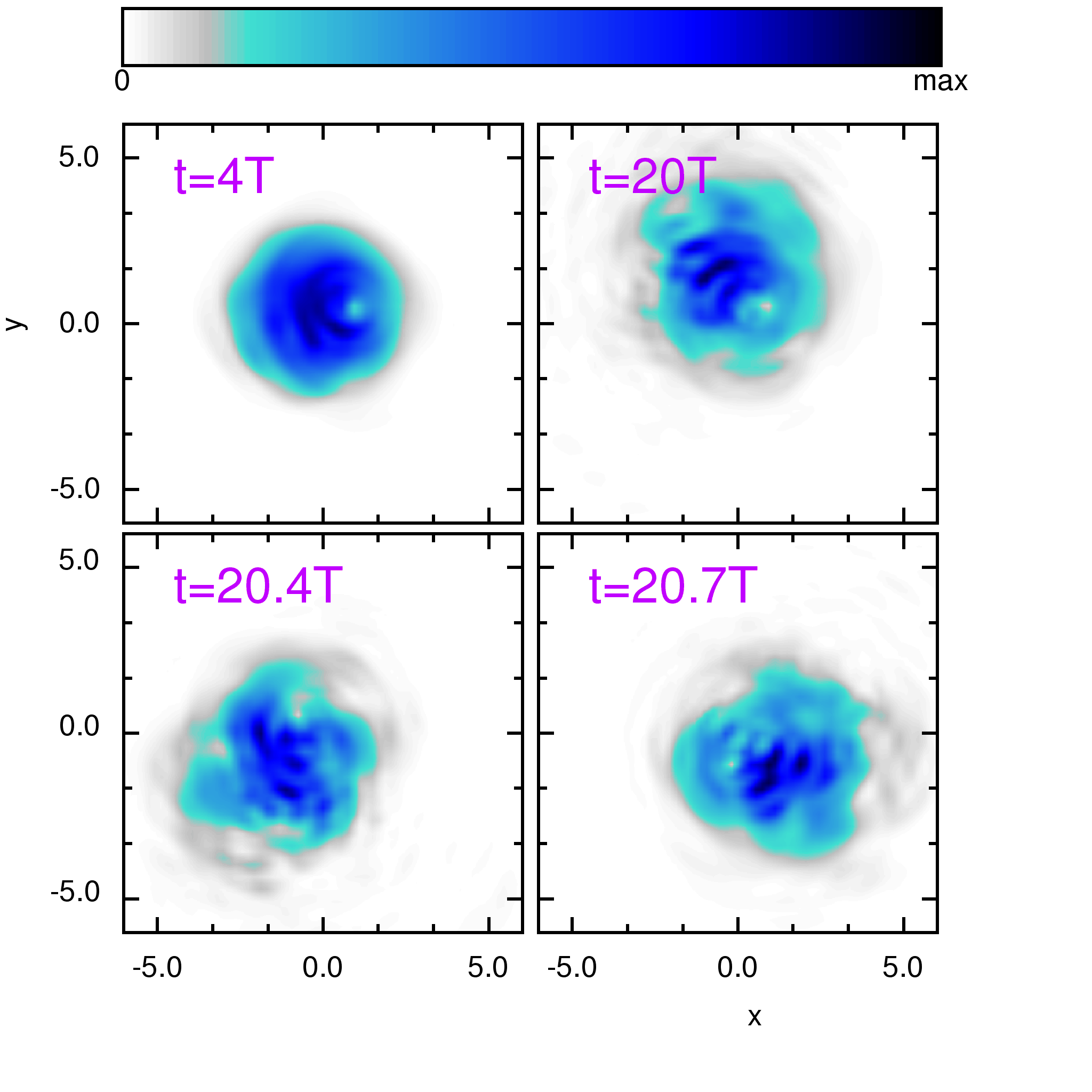}
\end{center}
\caption{(Color online) Resonant stirring of a 2D Bose gas at frequency $\Omega=0.986$. Plotted are screenshots of the density $\rho(r)$ at times $t=4T, t=20.1T, t=20.4T$ and $t=20.7T$, where $T=\frac{2\pi}{\Omega}$ is the period of one rotation. The gas first absorbs angular momentum by performing centre-of-mass rotation. After some certain time, vortices are formed in the bulk (second panel). The gas is in the highest resonance with respect to $\Omega$ [see also Fig~\eqref{Fig_EnergiesOm}]. The energy (it triples by $t=20T$) is absorbed both by the centre-of-mass rotation and vortex nucleation. The apparent node in the first panel is due to the rotating potential $V_{\text{rot}}$ of Eq.~\eqref{potentials}. The times plotted are chosen so that the centre-of-mass rotation of the gas is visible (note off-axes placement in the last three panels).}
\label{Fig_Res_Vor-Om986}
\end{figure}

\begin{figure}
\begin{center}
\includegraphics[width=0.75\linewidth,keepaspectratio]{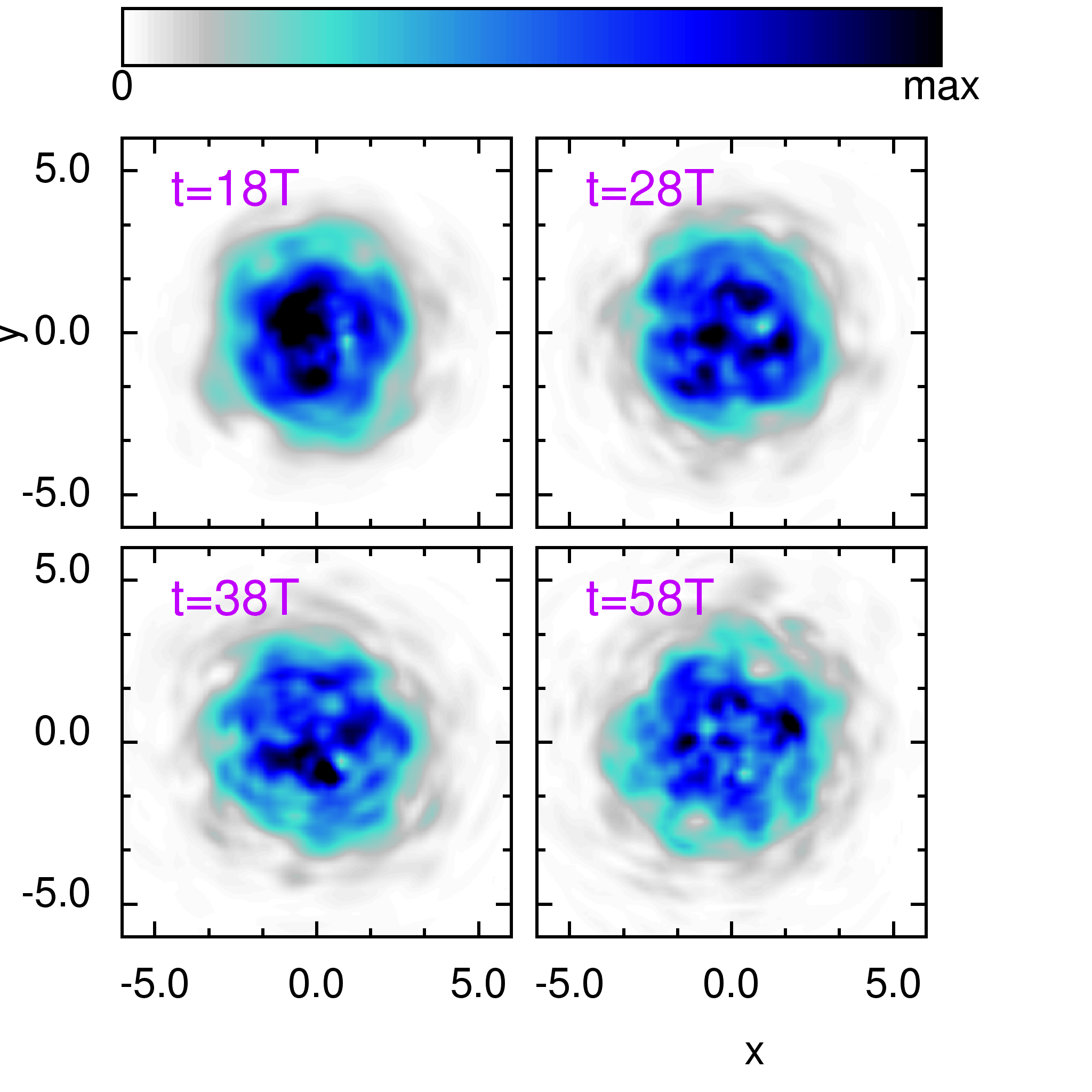}
\end{center}
\caption{(Color online) Resonant stirring with vortex nucleation at $\Omega=1.24$. The density $\rho(r)$ is plotted at times $t=18T$, $t=28T$, $t=38T$ and $t=58T$, where $T=\frac{2\pi}{\Omega}$ is the stirring period. Vortices are nucleated in the gas and their number increases in time. The fragmentation entropy $f$ reaches a maximum value already at $t=38T$ and the occupations reach the values $n_1=n_2=n_3=N/3$. The total energy of the system increases in time, apparently with no upper bound, for the whole range of our simulation. Note the total expansion of the gas, as compared to that of Fig.~\ref{Fig_Res_Vor-Om986}}. 
\label{Fig_Res_Vor-Om124}
\end{figure}

\paragraph{Resonant stirring with vortex nucleation.}
The highest energy absorption happens at $\Omega=0.986$. In this case a variety of excitations in the bulk of the gas appear, as can be seen in Fig.~\ref{Fig_Res_Vor-Om986}: vortex nucleation, expansion and centre-of-mass rotation. Vorticity is carried by the first fragment [orbital $\phi_1(\vec{r};t)$], and solitonic forms appear in the other two ($\phi_2$ and $\phi_3$). The fragmentation entropy $f$ reaches a maximum value at already $t=38T$, when the occupations take the values $n_1=n_2=n_3=N/3$ and stay such for the rest of the dynamics. The total energy seems to increase monotonically without bound, at least for the range of our calculations. The second most pronounced resonance for $t=22T$ occurs at $\Omega=1.24$ (see Fig.~\ref{Fig_EnergiesOm}). For this value we also plot the densities at different times in Fig.~\ref{Fig_Res_Vor-Om124}. In that case, vortex nucleation is seen, with no centre-of-mass rotation. We note that the spatially symmetric expansion of the bulk of the gas is also associated with energy absorption (compare the phenomenal sizes of the gas of Fig.~\ref{Fig_Res_Vor-Om124} to that of Fig.~\ref{Fig_Res_Vor-Om986}).

\begin{figure}
\begin{center}
\includegraphics[width=0.75\linewidth,keepaspectratio]{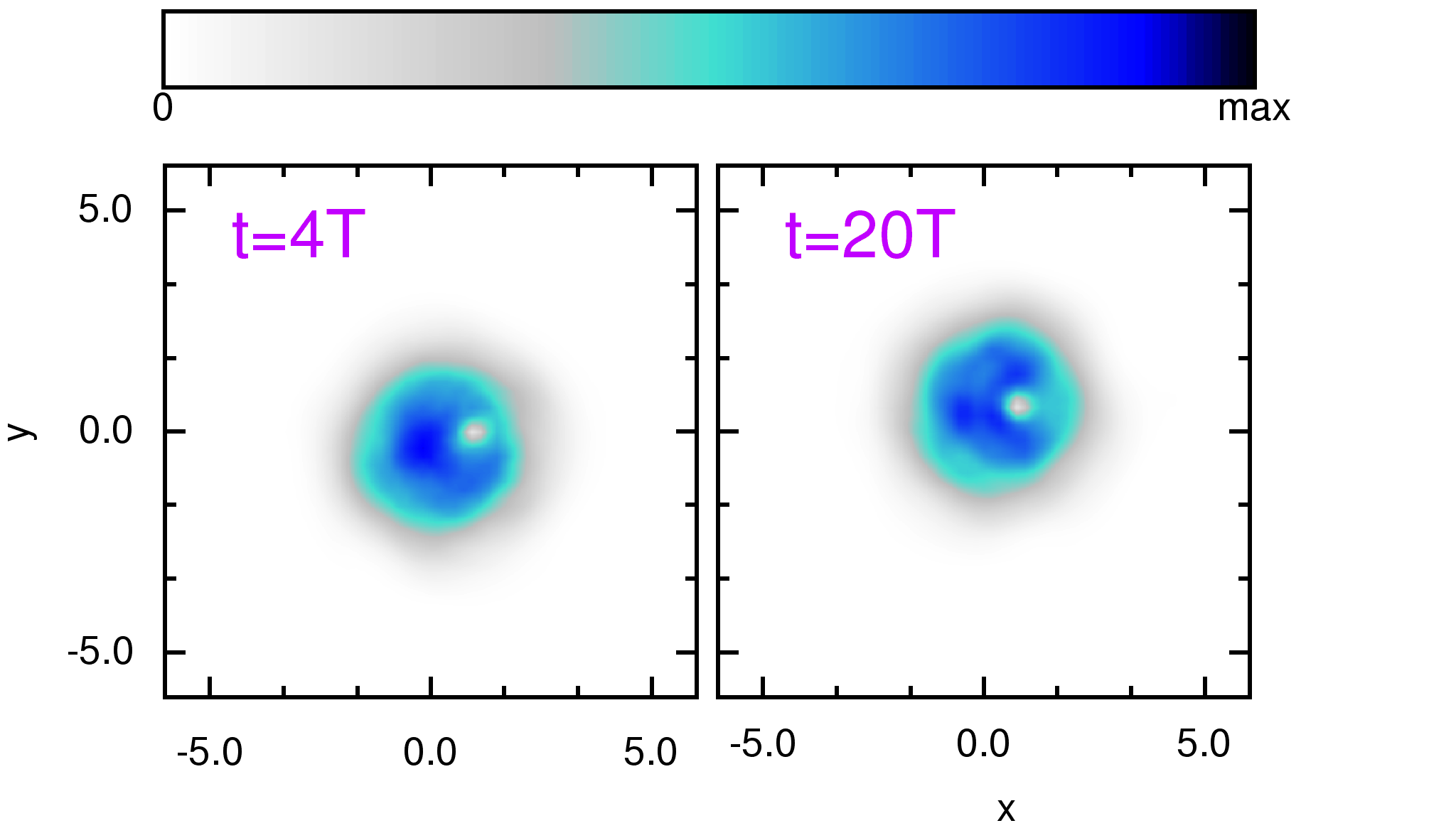}
\end{center}
\caption{(Color online) Off-resonant stirring at $\Omega=1.028$. Plotted is the density $\rho(r)$ of the system at times $t=18T$, and $t=30T$, where $T=\frac{2\pi}{\Omega}$ is the period of one rotation. Even though a slight off-axis motion of the gas can be seen, the density remains largely unchanged; it absorbs almost no energy from its environment and no vortices appear. Fragmentation is also close to zero throughout the dynamics. The apparent node in both panels is due to the rotating potential $V_{\text{rot}}$ of Eq.~\eqref{potentials}.}
\label{Fig_OffRes_NoVor-Om102}
\end{figure}

\paragraph{Off-resonant stirring.}
For the sake of completeness, we present the off-resonant case of $\Omega=1.028$. In Fig.~\ref{Fig_OffRes_NoVor-Om102} we plot the density of the gas. It can be seen that even after long stirring times the gas absorbs almost no energy from its environment and the fragmentation remains close to zero (see also Fig.~\ref{Fig_EnergiesOm}). The shape of the orbitals (and hence density) remains practically unchanged with respect to that at $t=0$.

\section{Discussion}
In the present work we have examined the time-evolution of the wavefunction of the 2D Bose gas that is continuously stirred with a rotating Gaussian laser beam. We saw a series of resonances as the stirring frequency increases, the most pronounced of which is at $\frac{\Omega}{\omega_{\text{trap}}}=0.986$. At this frequency the system appears to absorb energy from the environment (external rotor) at the maximum rate and fragments quickly. The fragmentation entropy achieves its maximum value for this resonance in the shortest time. 
The vortex nucleation implies that the system passes through states of different symmetry -- from zero to finite angular momentum -- and topology -- from zero to a finite number of nodes in the density. We conclude that the change in symmetry is accompanied by the emergence of fragmentation. As the system is stirred, it macroscopically  occupies excited orbitals. Slightly away from this resonant frequency (for instance $\Omega=1.02$) the system does not absorb energy and the total energy oscillates in time. The fragmentation in that case is close to zero and the density performs only a rigid-body-like centre-of-mass rotation.

We also identified secondary resonances and we examined the cases of $\Omega=1.00, 1.14$, interestingly, without vortices. In both cases the system fragments rapidly. However vortices are formed for some stirring frequencies. For instance, rotating at $\Omega=1.24$ results in about $5$ vortices being formed in the gas, after $t=38T$.

In conclusion, fragmentation appears and increases rapidly whenever there is resonances rotation and eventually vortex nucleation. This could hence invalidate a discussion on vortex nucleation within the Gross-Pitaevskii approximation, as it is not capable of capturing fragmentation effects (see also the broader discussion in Ref.\cite{Dagnino2009}). More generally, fragmentation becomes important in the description of out-of-equilibrium and strongly excited systems, where superfluidy breaks down (see for instance Ref.\cite{Yukalov2014}).

Last, we mention that the present work is part of an ongoing programme to investigate the manifestation of many-body phenomena, beyond the mean-field approximation in rotating and disturbed quantum gases with experimentally detectable signatures. The current analysis is not extensive and each of the briefly described excitation modes are to be further scrutinized. The here applied MCTDHB method is a powerful tool that gives information on the fragments (orbitals) of the state and their dynamics. However, one could look at the linear-response spectrum\cite{Grond2013} of the stationary many-body Hamiltonian in the corotating frame. This alternative approach would potentially identify the frequencies at which particular phases (centre-of-mass, vortex etc.) are excited retaining the benefits of an analytic treatment.

\begin{acknowledgements}
We thank V.S. Bagnato and S. Weiner for essential and useful comments on the manuscript. M.C.T acknowledges financial support from FAPESP. A.U.J.L. acknowledges financial support by the Swiss SNF and the NCCR Quantum Science and Technology. Computational time in the Hermit Cray computer of the HLRS is also gratefully acknowledged. Last, A.U.J.L. thanks Centro de Pesquisas em \'O{}ptica e Fot\^onica (CEPOF) of the IFSC of USP for generous hospitality. 
\end{acknowledgements}


\end{document}